\documentstyle[12pt]{article}
\begin{document}
\title{Non Commutativity, Fluctuations and Unification}
\author{B.G.Sidharth \\
B.M.Birla Science Centre,Hyderabad, 500463,India}
\date{}
\maketitle
\begin{abstract}
We point out that in the non commutativity and breakdown of
conventional spacetime at micro scales lies the seed to the unification of
gravitation and electromagnetism.
\end{abstract}
\section{Introduction}
Feynman had speculated that gravitation may be a result of some type of a
fluctuation, motivated by the fact that the attractive Van der Waal force can
be attributed to the dipole moments induced by the fluctuation in molecular
distribution\cite{r1,r2}. On the other hand the author had pointed out that
in a fluctuational scheme, not only do we get a key to electromagnetism, but
also we are able to deduce a consistent cosmology in which supposedly
accidental large number relations including the so called Weinberg relation
between the Hubble constant and the pion mass, arise naturally, as a consequence
of the theory\cite{r3,r4,r5,r6}. The question is, can gravitation also be
included in such a scheme, and if so can this point to the long sought after
unification of gravitation and electromagnetism?\\
We will now argue that indeed this is so - it is an underlying non commutativity
of the spacetime that has kept apart gravitation which relies on a smooth
spacetime manifold on the one hand, and Quantum Theory on the other hand,
though the latter also uses a smooth manifold as an approximation.\\
Relatively recent work by scholars like Ord, Nottale, El Naschie, the author
and others, particularly several papers in Chaos, Solitons and Fractals
have highlighted the non smooth and even fractal nature of spacetime\cite{r7,r8,r9,r10,r11}.
This again can be the key to Einstein's much sought after unification\cite{r12,r13,r14,r15}.\\
An understanding of the underlying substructure of spacetime is required for
such a unification program. Indeed Einstein himself had noted in the Journal
of the Franklin Institute,\cite{r16}"...it has ben pointed out that the
introduction of a space-time continuum may be considered as contrary to nature in
view of the molecular structure of everything which happens on a small
scale. It is maintained that perhaps the success of the Heisenberg method points
to a purely algebraic method of description of nature that is to the elimination
of continuous functions from physics. Then however, we must also give up, by
principle the space-time continuum. It is not unimaginable that human ingenuity
will some day find methods which will make it possible to proceed along such a
path. At present however, such a program looks like an attempt to breathe in
empty space". Wheeler\cite{r17} on the other hand notes, "the most evident
shortcoming of the geometrodynamic model as is stands is this, that it fails
to supply any completely natural place for spin $\frac{1}{2}$ in general and
for the neutrino in particular", while "it is impossible to accept any description
of elementary particles that does not have a place for spin half." Indeed he
describes the four dimensional spacetime as a classical approximation.
\section{Non Commutativity and Fluctuations}
We are beginning to realize now, and as we will also see in the sequel, that
the problem is that of reconciling the usual classical spacetime with the
spacetime of Quantum Theory, or as Witten puts it\cite{r18} Bosonic spacetime
with Fermionic spacetime.\\
Our starting point is the effect of an infinitessimal parallel displacement
on a vector\cite{r19}:
\begin{equation}
\delta a^\sigma = -\Gamma^\sigma_{\mu \nu} a^\mu d x^{\nu}\label{e1}
\end{equation}
Equation (\ref{e1}) represents the extra effect in displacements,
due to the curvature of space. In terms of partial derivatives with respect to the $\mu^{th}$
coordinate, (\ref{e1}) leads to,
\begin{equation}
\frac{\partial a^\sigma}{\partial x^\mu} \to \frac{\partial a^\sigma}{\partial x^\mu}
- \Gamma^\sigma_{\mu \nu} a^\nu\label{e2}
\end{equation}
where the $\Gamma$'s are the Christoffel symbols.
We consider the second term on the right side of (\ref{e2}) as:
$$-\Gamma^\lambda_{\mu \nu} g^\nu_\lambda a^\sigma = -\Gamma^\nu_{\mu \nu} a^\sigma$$
where we have utilized the linearity property that in the above formulation
$$g_{\mu \nu} = \eta_{\mu \nu} + h_{\mu \nu},$$
$\eta_{\mu \nu}$ being the Minkowski metric and $h_{\mu \nu}$ a small
correction whose square is neglected.\\
So from (\ref{e2}) we get,
\begin{equation}
\frac{\partial}{\partial x^\mu} \to \frac{\partial}{\partial x^\mu} -
\Gamma^\nu_{\mu \nu}\label{e3}
\end{equation}
From (\ref{e3}) we can deduce that
\begin{equation}
\frac{\partial}{\partial x^\lambda} \frac{\partial}{\partial x^\mu} - \frac{\partial}
{\partial x^\mu} \frac{\partial}{\partial x^\lambda} \to \frac{\partial}
{\partial x^\lambda} \Gamma^v_{\mu v} - \frac{\partial}{\partial x^\mu}
\Gamma^v_{\lambda v}\label{e4}
\end{equation}
If we now impose the condition that the right hand side in (\ref{e4}) does not vanish, then we have a non
commutativity of the momentum components in Quantum Theory. Indeed the left
side of (\ref{e4}) can be written as
\begin{equation}
\frac{1}{\hbar^2} [p_\lambda , p_\mu] \approx \frac{0(1)}{l^2}\label{e5}
\end{equation}
where $l$ is the Compton wavelength and $\hbar$ the reduced Planck length
wherein we have utilised the fact that at the extreme scale of the Compton
wavelength, the Planck scale being a special case, the momentum is $mc$.\\
We can, in view of (\ref{e3}) identify the right side of (\ref{e4}) as the
electromagnetic Field Tensor
$$\frac{e}{c\hbar}F^{\mu v}$$
What we have shown here is that once we consider the non commutativity of the
right or left side of (\ref{e4}), then it is meaningful to identify
\begin{equation}
A^\mu = \hbar \Gamma^{\mu v}_v\label{e6}
\end{equation}
with the electromagnetic four potential, thus leading to a unification of
electromagnetism with gravitation theory. This unification is not possible
in the usual commutative spacetime, that is when the right or left side of
(\ref{e4}) vanish. Equation (\ref{e6}), as we shall see can be deduced
alternatively.\\
Equation (\ref{e5}) is the manifestation of non commutative geometry: This has
been discussed in detail\cite{r15,r20}. The non commutative geometry is given by,
apart from (\ref{e5}), by relations like:
$$[x,y] = 0(l^2)$$
$$[x, p_x] = \imath \hbar [1+(a/\hbar)^2 p^2_x];$$
$$[t, p_t] = \imath \hbar [1-(a/ \hbar c)^2 p^2_t];$$
\begin{equation}
[x, p_y] = [y, p_x] = \imath \hbar (a/ \hbar)^2 p_xp_y ;\label{e7}
\end{equation}
$$[x, p_t] = c^2[p_{x,} t] = \imath \hbar (a/ \hbar)^2 p_xp_t ;\mbox{etc}.$$
Interestingly not only can the Dirac equation itself be deduced therefrom, but as
also the existence of the magnetic monopole at extreme Compton scales can be deduced.
Indeed this follows from (\ref{e5}) and the identification of the electromagnetic
Field Tensor: We have infact
$$Bl^2 \approx \frac{\hbar c}{e}$$
which is the celebrated monopole equation.
It must be mentioned that such strong magnetic fields can be shown to be
associated with a non commutative geometry from an alternative point of view\cite{r21}.\\
The relations (\ref{e5}) and (\ref{e7}) are a manifestation of the non point like,
spinorial structure of spacetime. In Quantum SuperString Theory,
Witten has called it
Fermionic spacetime, as against the usual commutative geometry of Classical or Bosonic
spacetime\cite{r18,r22}. This non commutativity is at the root of the
emergence of electromagnetism and spin as can be seen from (\ref{e5}) or
(\ref{e6}) and subsequent remarks.\\
Another way of looking at the non commutative relations (\ref{e5}) or
(\ref{e7}) is that, as pointed out by Witten, they provide a correction to the
Heisenberg Uncertainity Principle\cite{r23} viz.,
\begin{equation}
\Delta x \sim \alpha \frac{\Delta p}{\hbar}\label{e8}
\end{equation}
where, $\alpha \equiv l^2$, this as discussed in the references cited referring
to the well known duality, one encounters in Quantum SuperString Theory. In
this case $\Delta x$ is of the order of the radius of the universe and is
given by
\begin{equation}
\Delta x \sim \sqrt{N}l\label{e9}
\end{equation}
which is the well known Eddington Large Number formula, $N \sim 10^{80}$ being
the number of particles in the universe.\\
We now use the fact that the number of fluctuationally created particles is
$\sim \sqrt{N}$\cite{r3,r4}, so that the Uncertainity in the energy momentum is
given by
\begin{equation}
\Delta p \sim \sqrt{N}mc, \Delta E \sim \sqrt{N} mc^2\label{e10}
\end{equation}
Using equations (\ref{e8}), (\ref{e9}) and (\ref{e10}) along with equation
(\ref{e6}) we get, remembering that $h_{00}$ represents the gravitational
potential except for a factor $c^2$,
$$A^0 \sim \hbar \frac{\partial}{\partial x^0} h_{00} \sim \sqrt{N} mc^2
\left(\frac{Gm}{c^2}\right)$$
whence we get the celebrated large number "coincidence"
\begin{equation}
e^2/Gm^2 \sim \sqrt{N}\label{e11}
\end{equation}
Equation (\ref{e11}) provides the link between electromagnetism and gravitation
and arises from equation (\ref{e6}) and preceeding considerations with the
fluctuation input, very much in the spirit of Feynman. Infact the fluctuational
underpinning for interactions has been commented on earlier\cite{r24}.
\section{Remarks}
We observe that the identification of (\ref{e6})
with the electromagnetic four potential,leads to a unification of
electromagnetism with gravitation theory. This unification is not possible
in the usual commutative spacetime, that is when the right or left side of
(\ref{e4}) vanishes.\\
We now make a number of remarks which corroborate the
above deductions. The identification of (\ref{e6}) with the electromagnetic
vector potential was deduced and discussed at length though from a completely
different and infact Quantum Mechanical point of view\cite{r14,r3,r25}.
There the spinorial or pseudo vector
property of the Dirac four spinor was used, in a purely Quantum Mechanical
derivation.\\
This has been discussed at length in the references cited. But briefly, if the
Dirac bispinor is written as $\left( \begin{array}{ll}
\Theta \\ \phi
\end{array} \right)$, then at the Compton scale, it is the spinor
$\phi$ which predominates and moreover, under reflection,
$$\phi \to - \phi$$
This was shown to immediately lead to (\ref{e3}) or (\ref{e6}).\\
It was pointed out there that interestingly (\ref{e6}) was
mathematically similar to Weyl's original formulation except that Weyl had
introduced it ad hoc, infact as an external element, without any internal
derivation. This was why Weyl's formulation was rejected\cite{r26,r19}.
However as can be seen from the above, (\ref{e6}) is a consequence of the
pseudo spinorial behaviour at the Compton scale, which again is related to
the non commutativity of equation (\ref{e4}) and manifested in the non
commutative geometry contained in equations (\ref{e5}) and (\ref{e7}).\\
As string Theorist Greene puts it\cite{r27},"...wild electromagnetic field
oscillation, weak and strong free field fluctuations - quantum mechanical
uncertainity tells us the universe is a teeming, chaotic, frenzied arena on
microscopic scales...this frenzy is the obstacle to merging generaal relativity
and quantum mechanics." Only when we take into account the breakdown of the smooth
spacetime manifold do we begin to make progress.\\ \\
{\large {\bf APPENDIX}}\\ \\
In a recent paper El Naschie has introduced the concept of a fluction
\cite{r11}, a result of geometric fluctuation which could lead towards a
unification of fundamental forces. It is pointed out here that recent work by the
author does indeed emphasize the underpinning of fluctuations for fundamental
interactions.\\
In this recent work\cite{r25,r3,r28,r15,r29,r30},
it was pointed out firstly that the fluctuation of the electromagnetic field
(or the Zero Point Field) leads to\cite{r17},
\begin{equation}
\Delta B \sim \sqrt{\hbar c}/L^2,\label{e12}
\end{equation}
where $L$ is the spatial extent. It was pointed out that if $L \sim$ Compton wavelength
of a typical elementary particle then from (\ref{e12}) we recover the mass
and energy of this particle. In other words at the Compton wavelength the
elementary particle "condenses" out of the background Zero Point Field.
Similarly a fluctuation in the metric leads to (Cf.refs.\cite{r16,r25}),
\begin{equation}
\Delta \Gamma \sim \frac{\Delta g}{L} \sim l_P/L^2\label{e13}
\end{equation}
where $l_P \sim 10^{-33}cms \sim$ Planck scale. Unlike in equation (\ref{e12}),
if $L$ in (\ref{e13}) is taken to be $\sim l_P$ then from (\ref{e13}) we
get the gravitational interaction.\\
That fluctuations tie up equations (\ref{e12}) and (\ref{e13}) can be seen
explicitly as follows. As is known, given $N \sim 10^{80}$ elementary particles
in the universe, the fluctuation in the particle number is $\sim \sqrt{N}$ which
leads to a fluctuational electromagnetic energy which in the above scheme is
the energy of the typical elementary particle, so that we have (Cf. also\cite{r31})
\begin{equation}
\frac{e^2\sqrt{N}}{R} = mc^2\label{e14}
\end{equation}
Using in (\ref{e14}) the fact that\cite{r3,r4},
$$R = \frac{GNm}{c^2}$$
we get the well known relation
\begin{equation}
e^2 \sim Gm^2 \cdot \sqrt{N} = Gm^2 \cdot 10^{40}\label{e15}
\end{equation}
Equation (\ref{e15}) is usually interpreted as an adhoc or empirical relation
comparing the strengths of gravitational and electromagnetic forces. But once
the fluctuational underpinning has been taken into account, we have deduced (\ref{e15})
and can now see the connection
between electromagnetic and gravitational interactions. Indeed from (\ref{e15})
one can deduce that\cite{r32} at the Planck scale the electromagnetic and
gravitational forces become equal, or alternatively the Planck scale of mass $\sim 10^{-5} gms$
is a Schwarszchild black hole.\\
Indeed in the model referred to earlier, elementary particles like electrons
are Kerr-Newman type black holes giving at once both the electromagnetic and
gravitational fields including the Quantum Mechanical anomalous gyro magnetic
ratio $g = 2$\cite{r14}.\\
From this point, it was shown that the strong interactions follow at the Compton
wavelength scale itself, where the dimensionality is low (Cf.ref.\cite{r29,r28,r15}).
Infact within the same scheme, it was shown that the very puzzling characteristics
of quarks namely their fractional charge, handedness and confinement besides
the order of their massses can be deduced.\\
It is by the same argument of the fluctuation of the number of particles that
it was shown that the weak interactions can also be explained\cite{r30,r33}.
Indeed similar arguments in a different context were put forward years ago
by Hayakawa\cite{r31}.\\
Briefly if the weak force is mediated by a particle of mass $M$ and Compton
wavlength $L$ we get from the fluctuation of particle number, this time
$$g^2 \sqrt{N}L^2 \approx Mc^2 \sim 10^{-14},$$
whence the weak interaction can be characterised.\\
The conclusion is that the
spirit of fluctions is vindicated (Cf.also ref.\cite{r34}).


\begin{thebibliography}{99}
\bibitem{r1} R.P. Feynman, Lectures on Gravitation, Ed. B. Hatfield, Addison-Wesley,
New York, 1995.
\bibitem {r2} M.S. El Naschie, Chaos, Solitons \& Fractals, Vol.8, No.5,
1997, p.753-759.
\bibitem {r3} B.G. Sidharth, Int.J.Mod.Phys.A, 13 (15), 1998, p.2599ff.
\bibitem {r4} B.G. Sidharth, Int.J.Th.Phys., 37 (4), 1998, p.1307ff.
\bibitem {r5} B.G. Sidharth, Astronomy and Geophysics, Royal Astronomical
Society, London, April 1999, p.2.8.
\bibitem {r6} B.G. Sidharth, Nuovo Cimento, 115B (12,2), 2000, pp.151ff.
\bibitem {r7} B.G. Sidharth, Chaos, Solitons \& Fractals, 12(2001), p.173-178.
\bibitem {r8} B.G. Sidharth, "The Chaotic Universe", Nova Science Publishers,
New York, 2001 (in press).
\bibitem {r9} L. Nottale, Chaos, Solitons \& Fractals, (1994) 4, 3, p.361-388
and references therein.
\bibitem {r10} G.N. Ord, 04817 Elsevier Science CHAOS Ms 1036, MFC September
1998 (Chaos, Solitons and Fractals).
\bibitem {r11} M.S. El Naschie, Chaos, Solitons \& Fractals, 1999, 10(11), p.1947-1954.
\bibitem {r12} El Naschie, Chaos, Soliton \& Fractals, 12(2001), p.1361-1368.
\bibitem {r13} El Naschie, Chaos, Soliton \& Fractals, 12(2001), p.969-988.
\bibitem {r14} B.G. Sidharth, Gravitation and Cosmology, 4 (2) (14), 1998, p.158ff.
\bibitem {r15} B.G. Sidharth, Chaos, Solitons \& Fractals, 11(2000), p.1269-1278.
\bibitem {r16} M.S. El Naschie, Chaos, Solitons \& Fractals, 10(2/3), 1999, p.163.
\bibitem {r17} C.W. Misner, K.S. Thorne and J.A. Wheeler, "Gravitation",
W.H. Freeman, San Francisco, 1973, pp.819ff.
\bibitem {r18} W. Witten, Physics Today, April 1996, pp.24-30.
\bibitem {r19} P.G., Bergmann, "Introduction to the Theory of Relativity",
Prentice-Hall, New Delhi, 1969, p248ff.
\bibitem {r20} Y. Ne'eman, in Proceedings of the First Internatioinal Symposium,
"Frontiers of Fundmental Physics", Eds. B.G. Sidharth and A. Burinskii,
Universities Press, Hyderabad, 1999, pp.83ff.
\bibitem {r21} T. Saito, Gravitation and Cosmology, 6(2000), No.22, pp.130-136.
\bibitem {r22} B.G. Sidharth, "Quantum Super Strings and Quantized Fractal
Spacetime", to appear in Chaos, Solitons \& Fractals.
\bibitem {r23} B.G. Sidharth, Proceedings of Fourth International Symposium
on "Frontiers of Fundamental Physics" Kluwer Academy, New York (in press).
\bibitem {r24} B.G. Sidharth, Chaos, Solitons \& Fractals, 11(2000), p.2155-2156.
\bibitem {r25} B.G. Sidharth, Ind.J.Pure and Appl.Phys., Vol.35, July 1997,
pp.456-471.
\bibitem {r26} A. Einstein, "The Meaning of Relativity", Oxford \& IBH,
New Delhi, 1965, pp.93-94.
\bibitem {r27} B. Greene, "The Elegant Universe", Vintage, London, 2000, p.120.
\bibitem {r28} B.G. Sidharth, Mod.Phys.Lett.A., Vol.14, No.5, 1999, p.387-389.
\bibitem {r29} B.G. Sidharth, in Instantaneous Action at a Distance in
Modern Physics: "Pro and Contra" , Eds., A.E. Chubykalo et. al., Nova Science
Publishing, New York, 1999.
\bibitem {r30} B.G. Sidharth, Chaos, Solitons \& Fractals, 12(2001), p.1101-1109.
\bibitem {r31} S. Hayakawa,  Suppl of PTP Commemmorative Issue, 1965, 532-541.
\bibitem {r32} B.G. Sidharth, Chaos, Solitons \& Fractals, 12(2001), p.795-799.
\bibitem {r33} B.G. Sidharth, Chaos, Solitons \& Fractals, 12(2001), p.1449-1457.
\bibitem {r34} M.S. El Naschie, "Towards the Unification of Fundamental
Interactions....", to appear in Chaos Solitons and Fractals.
\end{thebibliography}
\end{document}